\documentclass{article}
\usepackage{graphicx}
\usepackage{amssymb, amsmath, amsfonts, amsthm, xypic}
\usepackage{color}

\usepackage{hyperref}
\usepackage{endnotes}
\let\footnote=\endnote

\theoremstyle{plain}% default

\theoremstyle{definition}

\theoremstyle{remark}

\newtheorem*{note}{Note}

\newcommand{\defeq}{\stackrel{\textrm{def}}{=}}
\newcommand{\lra}{\longrightarrow}

\newcommand{\M}{\mathcal{M}}

\newcommand{\A}{\mathbb{A}}
\newcommand{\Z}{\mathbb{Z}}
\newcommand{\R}{\mathbb{R}}
\newcommand{\C}{\mathbb{C}}
\newcommand{\Q}{\mathbb{Q}}

\newcommand{\K}{\mathbb{K}}
\newcommand{\T}{\mathbb{T}}

\newcommand{\CP}{\mathbb{CP}}
\newcommand{\Proj}{\mathbb{P}}

\newcommand{\Cat}{\mathcal{C}}

\DeclareMathOperator{\KO}{\textrm{KO}}
\DeclareMathOperator{\CH}{\textrm{CH}}
\DeclareMathOperator{\Spec}{\textrm{Spec}}
\DeclareMathOperator{\Sing}{\textrm{Sing}}
\DeclareMathOperator{\Vect}{\textrm{Vect}}
\DeclareMathOperator{\rank}{\textrm{rank}}
\DeclareMathOperator{\sing}{\textrm{sing}}
\DeclareMathOperator{\topo}{\textrm{top}}

\newcommand{\bnum}{\begin{enumerate}}
\newcommand{\enum}{\end{enumerate}}

\newcommand{\bitem}{\begin{itemize}}
\newcommand{\eitem}{\end{itemize}}

\newcommand{\bcenter}{\begin{center}}
\newcommand{\ecenter}{\end{center}}

\newcommand{\hlinen}{\begin{center}\line(1,0){250}\end{center}}

\date{}
\title{{\bf Dimension, Divergence and Desingularization}\footnote{A lightly-edited version of a submission to the 2012 FQXi Essay Contest.}\\ A Rough Cut on Motifs}
\author{Abhijnan Rej\footnote{Quantitative Finance Group, Tata Consultancy Services Ltd., 1 Software Layouts Unit, Madhapur, Hyderabad 500 081, Andhra Pradesh, India. Emails: {\tt abhijnan.rej@gmail.com} and {\tt abhijnan.rej@tcs.com}.}}
\begin{document}
\maketitle

\section{\underline{Das} Geometrie?}
\begin{figure}[h!]
  \centering
      \includegraphics[width=0.50\textwidth]{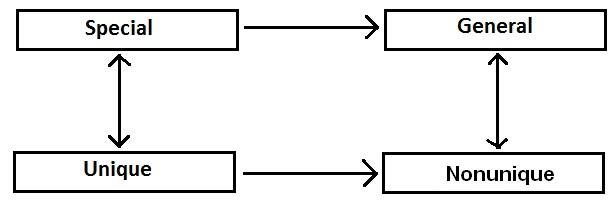}
  \caption{Uniqueness/nonuniqueness and the special/general}\label{pic:specialgeneral}
\end{figure}

The history of modern theoretical physics demonstrates a collective urge on behalf of its practitioners to seek \emph{beauty} and \emph{simplicity} in each generation's search for answers to, what it considers, fundamental questions about the material universe. In this desiderata one should also add the quest for \emph{uniqueness} of physical laws and an ontological one-to-one correspondence between ``law" and ``phenomena". Like the former, the notion of uniqueness of physical law is laden with subjective meaning-- an article of faith-- that struggles for a \emph{consistent} mathematical reformulation. At the same time, there is a tension between \emph{generality} and \emph{specificity} corresponding to nonuniqueness and uniqueness (see figure \ref{pic:specialgeneral}).
In this Essay, I examine this tension between uniqueness/nonuniqueness (and, \emph{inter alia}, generality/specificity) and mathematical consistency when it comes to a fundamental and foundational description of spacetime geometry coupled to matter in form of high-energy particles and forces in the form of the Standard Model\footnote{More precisely: I mean a manifold $(M,g)$ of dimension 4 and with a Lorentzian metric $g$ and with the pure gravity Lagrangian $R \sqrt{-\det g}$ coupled with the Standard Model Lagrangian, explicitly taking into account chiral fermions and flavor mixing of neutrinos (cf p. 166--167 of \cite{connes--marcolli-book} for an explicit expression.)}.
By looking at the issues of
\bnum
\item{the presence of divergences in quantum field theory, and}
\item{spacetime singularities,}
\enum
I will argue that 
\bcenter
\framebox{\parbox{15cm}{
consistent geometrical descriptions of the universe are \underline{far from unique} even as low-energy limits and that an abstract ``atomic" description of spacetime and gauge-theoretic geometry in terms of $K$-theories of algebraic and analytic cycles has the promise to restore uniqueness to our description. 
}}
\ecenter
\par
Early into the discovery of the Einstein equations of gravity in general relativity, it was found that even though the equations themselves were highly nonlinear, upon imposing certain classes of symmetries one could find a host of solutions, each describing a spacetime and, in some cases, its evolution. Some of them, like the G\"odel solution, was intriguing at first sight. (In case of the G\"odel solution, it described a universe with closed timelike curves-- a feature demonstrating the possibility of time travel.) It was only in the sixties and seventies that work on global causal structures by Penrose, Geroch, Hawking and others definitively identified, in an axiomatic way, which spacetimes are admissible as possible physical descriptions of our universe \cite{caus-struc}. 
\par
For Einstein and others, a logical next step after creation of general relativity was to seek out a classical geometrical unification of gravity with electromagnetism. For example, in a Kaluza--Klein model, the spacetime was taken as $\M \times S^1$ where $\M$ was a spacetime manifold of dimension 4 and $S^1$ a circle; the addition of the compact space $S^1$ was supposed to be a way to incorporate electrical charges into the theory. With hindsight we now know that even with the issue of quantization out of the way, such classical unification models can not exist in a meaningful way-- for example, some of such models renders the hydrogen atom unstable. Coupled with the plethora of such (physically indistinguishable) models, classical ``unified field theory"-- a dream of Einstein's literally till his dying day-- failed on both counts of consistency and uniqueness.
\par
The project of unification was more seriously taken up as a way to unify gravity not only with electromagnetism suitably quantized (in the theory of quantum electrodynamics) but with other two subnuclear forces (weak and strong interactions). A related Problem (with an uppercase `P') of quantizing the Einstein equations themselves remain unsolved till date. An unlikely push came from early research into hadronic interactions in form of string theory. As it stands in its supersymmetric version, string theory provides a glimpse on how the issue of unification of fundamental forces could eventually be resolved, as a problem in complex algebraic and differential geometry of 10-dimensional manifolds $\M \times C$ where $C$ is a compact manifold with some added geometrical structure\footnote{Typically, in ordinary type IIA or IIB superstring theory, $C$ is taken to be a Calabi-Yau manifold though in M-theory (which is 11-dimensional since it includes supergravity), $C$ can be as exotic as a \emph{$G_2$-manifold} which is a 7-dimensional Riemannian manifold with the automorphisms of the octonions as the holonomy group \cite{spiro}. Even within type IIA/IIA models, $C$ can also be an \emph{orbifold} which is, roughly speaking, a space with quotient singularities.}. 
\par
Unfortunately the scepter of nonuniqueness also haunts this program. Even with technical successes such as moduli stabilization (the KKLT formalism), it now seems plausible that instead of \underline{the} string theory, there are about $10^{500}$ solutions of something that should deserve that epithet or its 11-dimensional generalization incorporating supergravity (mystifyingly called M-theory.) While some cosmologists have whole-hearted taken in this ensemble as a physically-realistic setting through the mechanics of eternal inflation by viewing the universe as a patchwork of $10^{500}$ causally-disconnected regions of spacetime (called the Landscape), the worldview that it presents is one of extreme nonuniqueness. To spell it out, each of the solutions can be described using a different complex-geometric and topological datum and it is not at all clear, \emph{sans} appealing to some version of the Anthropic Principle, which is our neck of the woods. 

\section{Dimension and regularization}
At the heart of the claimed problem is how one distinguishes one geometrical space from another. In algebraic topology this is achieved by looking at various invariants (which are, conventionally, integers) coming from various groups attached to the spaces in question such as the topological Euler characteristic. A much more coarser number attached to a space is its \emph{dimension}-- a seemingly foolish way of distinguishing spaces $X$ and $Y$ for example might be in asking whether $\dim(X) = \dim(Y)$\footnote{In fact for basic algebraic objects such as vector spaces, it is enough to say that their dimensions are finite; this almost automatically makes them isomorphic.}. A first basic fact is that the 
\bcenter
\framebox{dimension of a space is a local notion.}
\ecenter
An ant roaming around on a donut can, in principle, conclude that the dimension of the donut (a torus, that is) is two whereas it will never be able to tell the genus of a donut to be one. More formally, a manifold $\mathcal{M}$ is of dimension $n$ if each arbitrarily small part of it around any arbitrary point looks like the Euclidean space $\R^n$. This, atleast, is the ``level-0" story and as with many other deceptively simple notions, one can say much much more.
\par
From Aristotle to Netwon the dimension of space was three. It was Minkowski who suggested that Einstein's relativistic geometry be best understood in terms of four dimensions. Most classical relativistic physics (unified field theory and such excluded) is defined on a four-dimensional spacetime. The formalism of relativistic perturbative quantum field theory (pQFT)-- renormalizing the theory by adding a finite number of counterterms to cancel ``bare infinities"-- crucially depend on the dimension of the spacetime it is defined on\footnote{For example, the $\phi^4$-scalar field theory is renormalizable only in dimension 4; a $\phi^6$-theory, on the other hand, is renormalizable only in dimension 6. However, generically speaking, a scalar field theory with potential $V(\phi) = \phi^k$ is \emph{not} renormalizable in $k$ dimensions!}. This is not a mere technical point. Contemporary particle since the seventies have used the criterion of renormalizability as a criterion of a \emph{permissible} physical theory. Therefore it makes sense to explore this further, through the well-known \emph{dimensional regularization} (DimReg) procedure. At the heart of this procedure is a formal definition involving a ``complex dimension" $D \in \C^\ast$ to extract finite values at $4 = D - \epsilon$:
\[ \int_{- \infty}^{\infty} e^{-\lambda k^2} d^Dk \defeq {\Big( \frac{\pi}{\lambda}\Big)}^{D/2}. \]
Through the work of Connes and Marcolli \cite{connes--marcolli-JGeomPhys, connes--marcolli-book}, we now understand the DimReg procedure in terms of the algebraic and noncommutative geometry of a space $X_\epsilon$ where $\epsilon \in \C^\ast$.
\par
Cutting a long story short (which links Feynman diagrams and the BPHZ preparation to Birkhoff factorization of loops on the Riemann sphere $\CP^1$ to flat equisingular connections and the Riemann--Hilbert correspondence and (!) finally onto \emph{motives}), as Connes has time and again reminded us, the main upshot of this picture is that
\bcenter
\fbox{[Connes--Marcolli] the renormalization group is a one-parameter subgroup of a motivic Galois group}
\ecenter
and that \cite{andre}-- from Galois' own terminology-- \emph{la th\'eorie de l'ambigu\"it\'e}.
\par
At the most elementary level, Galois' notion of ambiguity can be understood in the following way: suppose I have the following equation 
\begin{equation}\label{eq:quadratic} 
x^2 + 1 = 0.
\end{equation}
This equation clearly has no solution or \emph{root} in $\R$. On the other hand, over $\C$, (\ref{eq:quadratic}) has two roots $i$ and $-i$ and so the solutions of (\ref{eq:quadratic}) remain unchanged under the automorphisms
\[i \mapsto  i,\quad i  \mapsto -i.\]
(The ambiguity here is in the apparent lack of knowledge on which one of the two is \underline{the} solution to (\ref{eq:quadratic}).) Galois theory-- at the coarsest level of description-- encodes, for general degree polynomials and abstract fields, this ambiguity in terms of a subgroup of permutations of their roots (over a given field). 
\par
While Galois theory started with algebraic polynomials and their roots, viewed as a theory of ambiguity of solutions (of differential systems, \ldots) it extends to virtually every branch of mathematical theory (cf. section 3.1 of \cite{andre} for an example using solutions to a hypergeometric differential equation.) In the context of renormalization, Connes and Marcolli converted the analytic problem of dimensional regularization into a problem about fundamental groups of certain differential systems. Through the deep lens of Grothendieck--Galois correspondence, fundamental groups can also be viewed as Galois groups. Using this key insight, they should that the renormalization group is a Galois group of the form of a semidirect product
\[ \mathbb{U} \ltimes \mathbb{G}_m\]
where the group $\mathbb{U}$ is isomorphic to a very general object-- the \bcenter \fbox{motivic Galois group associated to mixed Tate motives over the scheme $\Spec \Z[i , \frac{1}{2}]$.}\ecenter The philosophical idea behind the work is the notion that the ambiguities obtained from scaling the coupling constants in renormalization can, in fact, be encoded in a group much like the ambiguities in the definition of roots of polynomial equations.
\par
However this turns out to be a double-edged sword; while it is certainly very nice to have a precise conceptual picture of renormalization, the form of the renormalization group thus identified is, in some sense, blind to the physical content of the theory. If the Connes--Marcolli picture is indeed \emph{physically} valid, it should give isomorphic renormalization groups for \emph{all} pQFTs that are renormalizable in the pertubative sense. This seems too strong a condition to impose, especially in the context of whether gravity is renormalizable perturbatively.
\bcenter
\framebox{\parbox{15cm}{
\underline{{\bf Sources of nonuniqueness \# 1}}
\\
The renormalization group is independent of the pQFT on which it acts on. In particular, it is independent of the matter content of the pQFT (presence of fermions, chirality, \ldots).
}}
\ecenter

\section{Spacetime singularities and their resolution}
Early into the discovery of general relativity, it was noticed that Einstein's equations admit spacetimes that are singular, i.e., there are points in the spacetime manifold $p \in \M$ in the neighborhood of which the curvature tensor on $\M$ diverges. While the first examples were shown to be pure artifacts of the choice of coordinates, it soon became clear that there were indeed genuine examples of singularities that can not be ``washed off" by coordinate transformations. Prominent examples of singularities include the ``center" of a blackhole, conical singularities (such as those arising in the study of gravitational lensing) and, most spectacularly, the Big Bang.
While the initial ``definitions" of these singularities were in terms of the explicit structure of the spacetime metric, it was structural definitions on the topology of $\M$ (Penrose, Geroch, Hawking, \ldots) that provides a mathematically sound description of spacetime singularities which is, simplifying a bit, the following, cf. \cite{caus-struc}:
\bcenter
\fbox{$\M$ is singular if $\M$ is geodesically incomplete.}
\ecenter
A prominent main argument for a quantum theory of gravitation is that quantum fluctuations of the metric $g_{\mu \nu}$ around the Planck scale $L_p \sim 10^{-35}$ meters will ``diffuse" the singularity; it was an early belief in string theory, that inverse string tension $\alpha' \sim L_p$ corrections to the spacetime curvature 
\[ R_{\mu\nu} + \frac{1}{2} \alpha' R_{\mu\rho\sigma \lambda} R_{\nu}^{\rho\sigma\lambda} + O({\alpha'}^2) \textrm{ terms} \]
would have smoothened out the (curvature) singularities. It was the important work of Horowitz and Steif \cite{horo-steif} that showed this \emph{not} to be the case-- classical singularities remain singular in a classical string background. Even in the first quantized case, they showed that string theory will not be able to eliminate these singularities.
\par
In the path integral formulation of quantum gravity, the Hawking--Hartle \emph{no-boundary proposal} \cite{hawk-hartle} replaces the singularity at the Big Bang with a smooth compact 3-surface with Euclidean signature\footnote{Hawking and Hartle define a ``wave function of the universe" which is a functional on the space of all compact 3-geometries and matter fields on these geometries. This wave function satisfies the Wheeler-DeWitt equation and the ground state amplitude on a prescribed 3-geometry is given by a path integral over all compact Euclidean 4-geometries which contain the prescribed 3-geometry as a boundary \cite{hawk-hartle}.}. While this is a tantalizing solution to the problem of the cosmological singularity, the main technical issue lies in the signature change of the spacetime manifold-- the transition of the Euclidean signature on the 3-surface to the Lorentzian one. However, as I will now argue, the Hawking--Hartle proposal has a nice analogue in algebraic geometry in the form of \emph{resolution of singularities} of algebraic varieties\footnote{An algebraic variety over (an algebraically closed) field $\K$ over dimension $n$ is, roughly speaking, the set of common zeros of a finite set of polynomials $f_i(x_1, \ldots, x_n)$ with coefficients in $\K$.}.
\par
To begin with, an algebraic variety $X$ is said to be \emph{singular} at a point $p \in X$ if $X$ is not flat in an infinitesimally small neighborhood of $p$. As we had earlier noted, since the dimension of a space is also a local notion, another way of saying $X$ is singular at $p$ is to say $\dim(X) \neq \dim(T_p(X))$, the dimension of the tangent space of $X$ at $p$. If $X$ is not singular at every $p \in X$, we say that $X$ is \emph{smooth}.
\par
\begin{figure}[h!]
  \centering
      \includegraphics[width=0.50\textwidth]{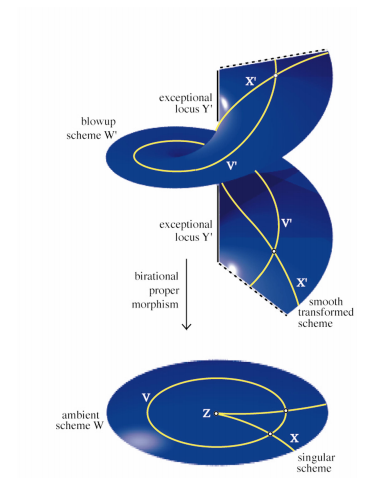}
  \caption{Blowing up $W$ to $W'$ (taken from p. 333 of \cite{hauser})}\label{pic:blowup}
\end{figure}
\begin{figure}[h!]
  \centering
      \includegraphics[width=0.50\textwidth]{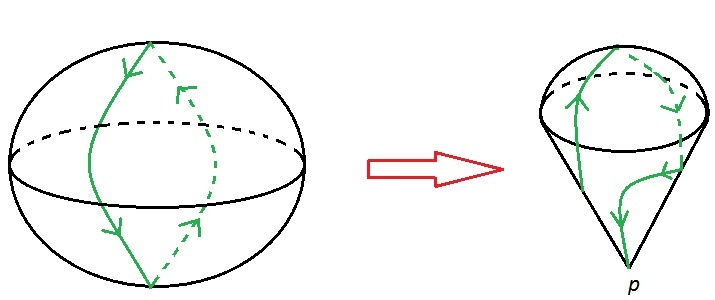}
  \caption{Geodesic incompleteness in an orbifold with an isolated singularity at $p$}\label{pic:orbifold}
\end{figure}
\begin{figure}[h!]
  \centering
      \includegraphics[width=0.70\textwidth]{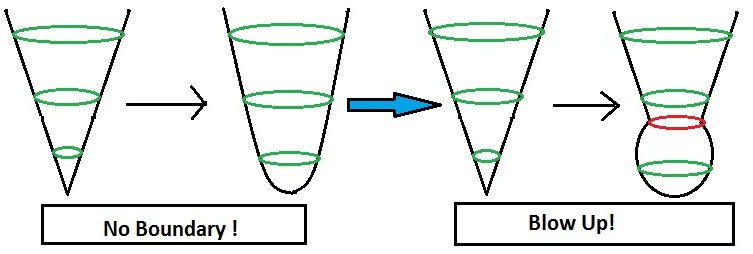}
  \caption{A comparison of the Hawking--Hartle no-boundary proposal (left) and the desingularization picture (right)}\label{pic:blownobound}
\end{figure}

This notion of  being ``locally not flat" can be extended to low-codimension spaces inside $X$ such as a line, surface, \ldots. Let us call such subspaces the \emph{singular sets} of $X$, denoting it as $\Sing(X)$. Roughly speaking, a resolution of singularities of $X$ is a ``well-behaved" map
\[ X \stackrel{\rho}{\longrightarrow} X'\] 
to a smooth variety $X'$ that induces an isomorphism\footnote{Technically $\rho$ is proper and birational.}
\[ X' \setminus \rho^{-1}(\Sing(X)) \simeq X \setminus \Sing(X).\] 
A typical way of resolving singularities is to take $\rho$ to be a \emph{blowup} of $X$ to $X'$ (though this alone is not sufficient!) Intuitively speaking, a blowup map disentangles the singular set in some higher dimensional space by spreading its image out. A picture being a thousand words, I refer the reader to figure \ref{pic:blowup} for an explicit example with the following set-up-- we consider the the real plane $W = \R^2$ (projectivizing it through the map $\R^2 \rightarrow \Proj^1$, $(x,y) \mapsto (x:y)$) with center $Z = \{0\}$. The space $V= \{x^2+y^2 = 1\}$ is a circle and $X$ is the singular curve $\{x^2 = y^3\}$ with a singularity $\Sing(X) = Z$. 
\bcenter
\framebox{\parbox{15cm}{
The surface $W' = \{xz -y = 0\} \subset \R^3$ in figure \ref{pic:blowup} is the blowup of $W = \R^2$ (after gluing the dotted lines).
}}
\ecenter
The main thing to note here is how the blowup map ``spreads out" the singular curve $X \subset W =  \R^2$ to $X'$ in $W' \subset \R^3$. It must be noted that it is not altogether clear whether one can always resolve low-codimension and nonisolated singularities of an algebraic variety or an analytic manifold. In fact, it is one of the great theorems of 20th century mathematics that
\bcenter
\framebox{\parbox{15cm}{
[Hironaka] Over a field of characteristic zero (such as $\R$ and $\C$), all singular varieties admit a resolution of their singularities.
}}
\ecenter 
How might this desingularization picture apply to physics? To begin with the context of pQFT, renormalization in coordinate space (and also, through the formalism of motives, in the momentum space) can be viewed as an exercise in the resolution of singularities of certain algebraic varieties naturally arising out of Feynman diagrams \cite{marcolli-book, ceyhan--marcolli}. For our purposes, a basic hint comes from the work \cite{balasub-etal} which presents a picture of a $(d+1)$-dimensional spacetime as an \emph{orbifold}. An orbifold, is roughly speaking, a manifold $X$ on which a group $G$ acts nonfreely-- the quotient of this manifold with the group action is called an orbifold $X/G$. The identification of points on $X$ after taking the quotient result in singularities in $X$, the number of isolated singularities being the number of fixed points of the group action $G \times X \rightarrow X$. So a main point to remember is that
\bcenter
\fbox{An orbifold $X/G$ is a singular space.}
\ecenter
Orbifolds are of particular interest in string theory where an admissible compactification is $\M \times X/G$ for suitable complex manifolds $X$ and suitable groups $G$. In fact one of the earliest example of a string target space was an orbifold $\T^6/\Z_3$. Blowing up $\T^6/\Z_3$ (which had 27 isolated singularities) gave a Calabi-Yau manifold \cite{aspinwall}. In  \cite{balasub-etal}, the authors identify the cosmological singularity with the orbifold $\R^{1,d}/\Z_2$. It has been well known for some time now that an orbifold can represent spacetime singularities well in the sense that geodesic incompleteness has a natural interpretation in case of orbifolds \cite{horo-steif} (see figure \ref{pic:orbifold}) so it makes sense to use them as models of singularities\footnote{In case of gravitational lensing by a cosmic string, the cosmic string is viewed as giving us a conical singularity which has a ready description as an orbifold with one singularity at the vertex of the cone.}. 
It goes without saying that orbifolds (as algebraic or analytic manifolds over $\C$) admits a resolution of singularities so viewing spacetime singularities as such objects give us a mathematically well-defined problem of resolution of their singularities. As an example, consider figure \ref{pic:blownobound}. 
Viewing spacetime as an orbifold leads to the possibility of resolving the singular boundary in $\R^{1,3}$ (the cosmological or Big Bang singularity) by blowing it up. This is, qualitatively, the same as replacing the singular point with a 2-sphere in an ambient $\R^3$ of figure \ref{pic:blownobound}. The result is expected to be geometrically different from the no-boundary prescription-- in particular, the physics of the picture is that of a universe expanding and then contracting and then expanding again (with a much bigger rate of expansion at the second stage)\footnote{There is a possibility that the ``blowup picture" of resolving the cosmological singularity in figure \ref{pic:blownobound} can be modeled using phases of the mixmaster universe of Belinskii, Khalatnikov, and Lifshitz.}. Modulo the important issue of the identification of the ``correct" group $G$ whose action of a smooth $\R^{1,3}$ leads to the orbifold $\R^{1,3}/G$ with $\Sing(\R^{1,3}/G)$ the singular spacetime boundary, this picture is rather attractive\footnote{There are important issues that I am completely ignoring here-- for example, the existence of a Big Bang singularity leads to specific restrictions on the causal structure of the spacetime. A much more serious description of the cosmological singularity in terms of orbifold singularities and their resolution has to take all of that into account.}-- for example, the discovery of a suitable $G$ might tells us \emph{why} we have cosmological singularities to begin with. Engaging in complete over-the-top speculation, such a group might even be related to symmetries of exotic (nonscalar) matter fields present at the Planck scale. 
\par
However, this description (if correct), leads to another source of nonuniqueness of the spacetime geometry stemming from the desingularization procedure itself. 
\par
First of all, topological (diffeomorphism) invariants of a space $X$ generally cease to be so after $X$ has been blown up. For example, the canonical class $K_X$ of an analytic manifold $X$ ceases to be the same after a blowup $X \rightarrow X'$, so it has been clear for some time now that we have to put additional restrictions on the blowups that we allow; in case of $K_X$, we would additionally require that the resolutions to be \emph{crepant}. This, by definition, means a blowup $X \rightarrow X'$ that gives $K_X = K_{X'}$. 
\par
Second of all, absent explicit invariants associated to the desingularization procedure, we do not know whether a certain desingularization $X \rightarrow X'$ is the ``optimal" in the sense that we performed the minimal number of blowups that leads to the desired desingularition datum\footnote{By the ``desired desingularization datum", I mean a smooth space $X'$ with an exceptional divisor $E$ in $X'$  that has normal crossings. \par Results of \cite{bier-mil} seems to establish a canonical desingularization procedure with an associated algorithm though the introduction of a new invariant in their work which seems to be unrelated to other known local geometric invariants gives the procedure an artificial feel.}.
\par
Therefore, to our list of geometric nonuniqueness we can add the following entry:
\bcenter
\framebox{\parbox{15cm}{
\underline{{\bf Sources of nonuniqueness \# 2}}
\\
A resolution of singularities of a space $X \rightarrow X'$ that either topologically models our universe or describes it as a low-energy limit is neither canonical nor does it preserve topological invariants. Topological invariants (such as the Chern class) explicity depends on the model of desingularization. 
}}
\ecenter
In fact, we can go and take a step back with the issue of spacetime singularities and ask whether we should take a singular (analytic/algebraic) space as our basic model of spacetime. As I have remarked earlier about the work \cite{horo-steif}, even granted string theory, our description of spacetime at the fundamental level is likely to contain singular points and boundaries where the description by local geodesics break down. If this is indeed the case, then we run into very deep issues about the natural existence of topological invariants, cf. \cite{totaro} for a quick summary of aspects of this problem most likely to be relevant for physics. 
\par
A basic first problem would be that in a space with even mild singularities, Poincar\'e duality (which pairs cohomology groups in complimentary degrees) fails. This is a very serious problem since (co)homology groups, in more than one sense, contain information about all possible subspaces of the given space, and an important application of Poincare duality is in the pairing of topological information about the ambient space and that of its subspaces\footnote{A very general formulation of Poincar\'e duality for smooth projective varieties over a field $\K$ would be the following. Let $X$ be such a variety and $r = \dim(X)$. Then
\[H^{2r}(X, \K) \simeq \K.\]
Equivalently, Poincar\'e duality would be a nondegenerate pairing
\[H^{2r - i}(X, \K) \times H^{i}(X, \K) \simeq \K.\]
}.
\par
\begin{figure}[h!]
  \centering
      \includegraphics[width=0.35\textwidth]{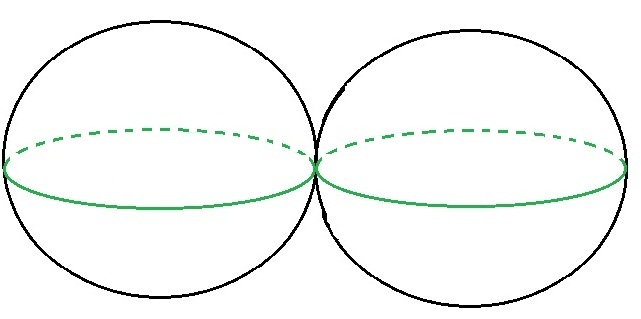}
  \caption{The singular curve $X$ in $\mathbb{CP}^2$}\label{pic:twospheres}
\end{figure}

As an example (taken from \cite{banagl--maxim}) consider the curve \[X = \{(x:y:z) \in \CP^2| xy = 0\}\] in figure \ref{pic:twospheres}.
This curve $X$ is just two copies of the Riemann sphere $\CP^1$ meeting at a single (singular) point $(0:0:1)$. Therefore topologically $X = S^2 \bigvee S^2$. Computing the cohomology groups we see that 
\[H_0(X, \Q) = \Q, \quad H_2(X, \Q) = \Q \oplus \Q\]
and therefore \[\rank(H_0(X, \Q)) \neq \rank(H_2(X, \Q))\]
which contradicts Poincar\'e duality. 
\par
In fact, almost all of the commonplace topological notions that we love and use in geometrical physics (Chern classes, Riemann--Roch theory, \ldots) fail to hold for singular spaces as they stand and have to be replaced by far more sophisticated notions such as intersection cohomology (cf. \cite{banagl--maxim} for a discussion geared towards mirror symmetry) and Chern--Schwartz--MacPherson theory. To reiterate this point,
\bcenter
\framebox{\parbox{15cm}{
\underline{{\bf Sources of nonuniqueness \# 3}}
\\
If the geometry of spacetime is singular at a basic level (that is, singular spactimes stay singular in any fundamental theory of quantum gravity) then our basic algebraic-topological tools have to be replaced by ``something else". In particular, the breakdown of Poincar\'e duality would imply a fundamental difference between computing cohomology groups and computing homology groups that have to reconciled in the more general theory.}}
\ecenter

\section{Coda-- cycles and ``atoms of space"}
To summarize and surmise the points so far-- we have seen that geometries of spacetime coupled to matter or even as models of pure gravity tend to possess highly nonunique features. In case of the geometry of renormalization (the Connes--Marcolli theory), the nonuniqueness lies in the establishment of the renormalization group as a very general mathematical object; whether this is a matter of virtue or not is clearly an contentious issue. Less contentious are the problems that arise whenever we try to establish a singular spacetime as a foundational notion. The problems with definitions of topological invariants in such a case (the failure of Poincar\'e duality being an example) and attendant problems with the desingularization of a such a spacetime (replacing the cosmological singularity by a blowup of the singular boundary of spacetime) clearly points toward replacing our ``more or less smooth" notions with far more refined tools. As a conclusion I will argue in this section that such tools already exist in the form of $K$-theory and its generalizations and ultimately in that other \emph{theory of everything}-- Grothendieck's notion of motives of spaces. (Immodestly I refer the reader to \cite{rej} for a ``physically relevant" summary of motives.)
\par
To begin with then, given a manifold $M$, what is the most important topological ``gadget" that can be attached to it with which gives us the most important invariants of $M$? Both in the algebraic and the analytic categories, the answer would be-- its cohomology ring $H^\ast_{-}(M, \K)$ with coefficients in a field $\K$. (The inexperienced reader should immediately replace this by $H^\ast_{\sing}(M, \Q)$.) 
\par
\begin{figure}[h!]
  \centering
      \includegraphics[width=0.25\textwidth]{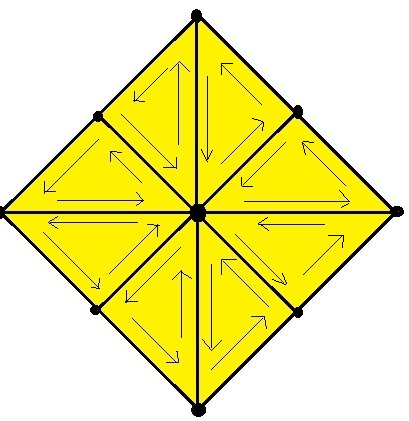}
  \caption{Cycles and 2-simplices in $\R^2$}\label{pic:newsimplexcycle}
\end{figure}

At the heart of the construction of a cohomology ring is the decomposition of the space in terms of cycles, chains and their complexes (see figure \ref{pic:newsimplexcycle} for a baby example). This is how we define singular homology, for example.
In the algebraic category, we work with an equivalent notion-- the Chow ring-- which is constructed out of the free abelian group of formal sums of irreducible subspaces on a nice variety $X$ and an appropriate equivalence relation between these formal sums (which we would call \emph{algebraic cycles}). This ring (graded by the \emph{co}dimension of the subspaces) encodes most of the topological invariants in the sense that there are ``functorial" maps from pieces of the Chow ring
\[\CH^i(X) \otimes \Q \lra H^{2i}(X, \Q).\]
It remains a very important problem in algebraic geometry (and perhaps, all of mathematics) to classify and understand the images of these maps for given varieties $X$.
\par
I have, so far, alluded to the cohomology ring containing important information about topological invariants. Let us delve a little deep to make more sense of the statement. Let $\mathsf{C}$ be a category of ``good" topological spaces (compact,with finite compact support, \ldots). The \emph{topological Euler characteristic} of an object in $\mathsf{C}$ is defined as
\[ \chi_{\topo}(X) \defeq \sum (-1)^k \dim H_c^k(X, \Q).\]
The topological Euler characteristic satisfies the inclusion-exclusion principle: let $Y \hookrightarrow X$ be a  closed subspace and $U = X \setminus Y$. Then
\[\chi_{\topo}(X) = \chi_{\topo}(Y) + \chi_{\topo}(U).\]
With this we construct a very general notion of a Grothendieck group of the category of good spaces $\mathsf{C}$:
\[ K_0(\mathsf{C}) \defeq \textrm{ abelian group generated by the relations }[X] = [Y] + [U].\]
Letting $K_0(\Vect_\Q)$ be the free abelian group of finite dimensional $\Q$-vector spaces modulo the relation
\[ V \sim U \oplus W  \iff V \oplus U = W,\]
we can view the topological Euler characteristic as a composition of maps (\cite{peters})
\[ \chi_{\topo}: K_0(\mathsf{C}) \longrightarrow K_0(\Vect_\Q) \stackrel{\sim}{\longrightarrow} \Z, \]
the last isomorphism being taking dimension of vector spaces $[V] \mapsto \dim_\Q(V)$.
\par
The main point here for the motivic object $K_0(\mathsf{C})$ is that
\bcenter
\framebox{\parbox{15cm}{
We can additively decompose any arbitrary good topological space $X$ as sums of ``elementary" units in $K_0(\mathsf{C})$.}}
\ecenter
The elementary units are $[\A^0]$, the class for a point which we will denote as 1, and $[\A^1]$ the class of a (affine) line $\A^1$ which we will denote as $\mathbb{L}$. For example, the projective $n$-space can be decomposed as
\begin{eqnarray*} [\Proj^n] =  
1 + \sum_{i = 0}^{n-1} \mathbb{L}^i
			& = &  \frac{1 - \mathbb{L}^{n+1}}{1- \mathbb{L}} \\
                        & = & \frac{{(1 + \mathbb{T})}^n - 1}{\mathbb{T}}
\end{eqnarray*}
where $\mathbb{T} = [\A^1] - [\A^0]$ is the class of the 1-torus.
\par
The Grothendieck group allows for ``Lego"-like decomposition of a space into ``atoms" arising out of algebraic cycles. In the special case of  mixed Tate motives-- a class of objects of intense current interest in the study of pQFT, cf. \cite{marcolli-book}-- the atoms are $1$ and $\mathbb{L}$; in other words, mixed Tate motives form a polynomial ring
\[ \Z[\mathbb{L}, {\mathbb{L}}^{-1}] \subset K_0(\mathsf{C}).\]

 In terms of resolution of singularities, it allows for explicit algebraic expressions that separate out the space that is being blown up, the subspace along which the blow up is being made as well as the exceptional divisor arising from the blow up. For example, let $X$ be a space being blown up along subspace $Y$ of codimension $d$  and denote the resulting space as $\textrm{Bl}_Y(X)$ and the exceptional divisor as $E$. One of the early results on motives from the sixties tells us that in $K_0(\mathsf{C})$
\[[\textrm{Bl}_Y(X)] = [X] + \mathbb{L}[E] - \mathbb{L}^d[Y].\]
With these simple examples at hand, I would argue that an abstract algebraic notion like $K_0(\mathsf{C})$ should be the proper ``premise" in which we look at the physical spacetime geometry and its ``atomic structure". Already motivic ideas have been incorporated in both loop quantum gravity (in terms of correspondences and spin networks) as well as by the use of algebraic cycles in mirror symmetry so it should not come as a surprise altogether that we replace our conventional-- and often ambiguous-- geometrical notions for something more deeper. 
\par
Perhaps the `M' in M-theory also means \emph{motivic}?
\hlinen

\newpage
\def\enotesize{\normalsize}
\theendnotes

\newpage
\bibliographystyle{amsalpha}

\begin{thebibliography}{99}

\bibitem{andre}
Y. Andr\'e, ``Ambiguity theory: old and new", arXiv: 0805.2568, 2008.

\bibitem{aspinwall}
P.S. Aspinwall, ``Resolution of orbifold singularities in string theory", \emph{Mirror Symmetry II} (B. Greene \& S.T. Yau, eds.), American Mathematical Society (2001), 355--379.

\bibitem{balasub-etal}
V. Balasubramaniam et.al., ``A space-time orbifold: a toy model for a cosmological singularity", \emph{Phys. Rev. D}, {\bf 67}, 026003 (2003).

\bibitem{banagl--maxim}
M. Banagl and L. Maxim, ``Intersection spaces and hypersurface singularities", preprint, May 2012.

\bibitem{bier-mil}
E. Bierstone and P.D. Milman, ``Resolution of singularities", \emph{Several Complex Variables}, MSRI Publications {\bf 37}, MSRI Berkeley (1999), 43--78.

\bibitem{ceyhan--marcolli}
O. Ceyhan and M. Marcolli, ``Feynman integrals and motives of configuration spaces" \emph{Communications in Mathematical Physics},  {\bf 313},  no.1 (2012), 35--70.
\\
O. Ceyhan and M. Marcolli, ``Feynman integrals and periods in configuration spaces", arXiv: 1207.3544, 2012.

\bibitem{connes--marcolli-JGeomPhys}
A. Connes and M. Marcolli, ``Quantum fields and motives", \emph{Journal of Geometry and Physics}, {\bf 56}, no. 1 (2005), 55--85.

\bibitem{connes--marcolli-book}
A. Connes and M. Marcolli, \emph{Noncommutative Geometry, Quantum Fields and Motives}, Colloquium Publications Vol.55, American Mathematical Society, 2008.

\bibitem{hauser}
H. Hauser, ``The Hironaka theorem on resolution of singularities", \emph{Bulletin of the AMS}, {\bf 40}, no. 3. (2003), 323--403.

\bibitem{caus-struc}
S.W. Hawking and G.F.R. Ellis, \emph{The Large Scale Structure of Space-time}, Cambridge Monographs in Mathematical Physics, Cambridge University Press, 1975.

\bibitem{hawk-hartle}
S.W. Hawking and J.B. Hartle, ``Wave function of the universe", \emph{Phys. Rev. D.}, {\bf 28}, no. 12 (1983), 2960--2975.

\bibitem{horo-steif}
G. T. Horowitz and A.R. Steif, ``Spacetime singularities in string theory", \emph{Phys. Rev. Lett.}, {\bf 64}, (1990), 260--263. 

\bibitem{spiro}
S. Karigiannis, ``What is a $G_2$-manifold?", \emph{Notices of the AMS}, {\bf 58}, no. 4. (2011), 580--581. 

\bibitem{marcolli-book}
M. Marcolli, \emph{Feynman Motives}, World Scientific, 2010.

\bibitem{peters}
C. Peters, \emph{Motivic Aspects of Hodge Theory}, Narosa Publishing House/Tata Institute of Fundamental Research, 2010.

\bibitem{rej}
A. Rej, ``Motives: an introductory survey for physicists" (with an appendix by Matilde Marcolli), \emph{Contemp. Math.},  {\bf 539}, American Mathematical Society (2011), 377--415.

\bibitem{totaro}
B. Totaro, ``Topological invariants of singular spaces", lecture at University of Michigan, Ann Arbor, April 18, 2005. Notes by S. Payne.







\end{thebibliography}

\end{document}